\documentclass{revtex4}
\usepackage{graphicx}

\def\head#1{\null\vskip -25pt\null \section{#1} \vskip -05pt}

\begin{document}
\bibliographystyle{revtex}

\setlength{\textheight}{241mm}
\setlength{\textwidth}{170mm}

\def\figiii{
\begin{figure}[htb]
\epsfxsize=0.45\hsize \epsfbox{z.eps}
\epsfxsize=0.45\hsize \epsfbox{rap.eps}
\caption{Binned distribution in $z$ (left: $\eta$ integrated out)
and $\eta$ (right: $z$ integrated out), each for a fine and a broad 
binning. 
\label{fig:1bin}
}
\end{figure}
}

\def\eq#1{Eq.~(\ref{eq:#1})}
\def\fig#1{Fig.~(\ref{fig:#1})}
\def\pt{p_T}
\def\sec#1{Section~\ref{sec:#1}}

\preprint{
\hfill{\small MSUHEP-?????} \\
\hfill{\small hep-ph/0112307}
}


\title{PDF's: What Do We Need To Know?}
\thanks{
Contributed to  
E1-C: Retro/Exotic Neutrino Interactions subgroup; 
APS/DPF/DPB Summer Study on the Future of 
Particle Physics (Snowmass 2001), 
Snowmass, Colorado, 30 June - 21 July 2001.
}

\author{Fredrick Olness}
\email[email: ]{olness@mail.physics.smu.edu}
\homepage[   ]{http://www.physics.smu.edu/~olness}
\affiliation{
Department of Physics,  Southern Methodist University, Dallas, TX 75275-0175
}

\date{\today}

\begin{abstract}
\noindent

Reliable knowledge of parton distribution functions (PDFs) is crucial
for many searches for new physics signals in the next generation of
experiments.  Presently, there remain a number of open questions
regarding the PDF's and their uncertainties.  We briefly discuss these
issues, and consider aspects where a future high-precision
fixed-target experiment might contribute.\cite{note}

\end{abstract}

\maketitle

\head{Introduction} \label{sec:intro}

In the past few years there has been considerable progress towards
understanding some of the uncertainties in the individual measurements that
contribute to our knowledge of parton 
distributions (PDFs).\cite{Lai:2000wy,Giele:2001mr}
 While high energy collider experiments can cover a large kinematic 
region in $\{x,Q^2 \}$-space, there are still a number of questions
that can be best answered by a high-precision low-energy fixed-target 
experiment.

\head{Nuclear Effects} \label{sec:}

To decipher the flavor content of the proton, it is essential to use
both charged and neutral current processes.  For the charged current
process, neutrino DIS plays an important role as the basic process
$\nu N \to \ell^{\pm} X$ contains an easily detected charged lepton
($\ell^{\pm}$) in the final state.  Unfortunately, as the neutrino 
cross section is relatively small, we are forced to use massive
targets with high-Z nucleons in order to obtain reasonable statistics.
 For example, in the case of NuTeV, an iron (Fe) target was used.
Scaling the $\nu$-Fe cross section to a $\nu$-isoscalar cross
section poses a significant challenge as this involves complex nuclear
effects for which we do not have a comprehensive theory.

When we attempt to extract PDF's from $\nu$-DIS on heavy nucleons our
task is compounded by the difficulty that we must simultaneously deal
with these nuclear effects and the separation of the PDF 
flavors.\cite{Brock:1995sz}
 Clearly, this is an area where a future high luminosity neutrino
factory could considerably improve our understanding of the $\nu$-DIS
process.\cite{Albright:2000xi,Mangano:2001mj,Bigi:2001xb} 
Such a facility would allow us to perform high statistics
measurements on light targets (hydrogen, deuterium, ...)  thereby
enabling us to separate the nuclear effects from the nucleon
structure.\cite{Bock:1997wi}
 
Specifically, one can envision a comprehensive two-step program.
 1) Use light targets to accurately determine the 
flavor structure of the proton by combining this data with 
neutral current data also taken on light targets. 
 2) Compare the light target data (both charged and neutral current processes)
with data from heavy targets to extract the nuclear effects. 
 Such a program would represent a tremendous advance in the understanding
of both the proton structure and the corrections due to high-$Z$ nuclei.

\head{PDFs at Large-$x$} \label{sec:}

While nuclear binding effects are clearly seen in heavier nuclei, the
size of these effects for Deuterium is under debate.
 Specifically, the ratio of the density of down quarks to that of up
quarks in the proton in the region $x\rightarrow1$ can provide
valuable information on both higher twist corrections and nuclear
binding effects.  This issue is complicated as: 1) it depends
critically on a nuclear physics model with many parameters fit to the
data, and 2) the deuteron is a very special nucleus with binding
energies much smaller than the rest, so that a large extrapolation
from the heavier nuclei is needed.

While future high luminosity HERA measurements of positron-induced
charged current interactions will address this issue, preliminary
calculations indicate a very large luminosity is required to impose
strong constraints on possible nuclear corrections.
 Conversely, one possible way to constrain the $d$ quark is from
measurements of $\pi^{+} /\pi^{-}$ production in DIS 
interactions.\cite{Kuhlmann:2000sf} 
A careful design of the beam and detector would allow for precision
measurements in the large-$x$ region that are competitive with
collider facilities.

\head{Extraction of $s(x,Q^2)$ from neutrino DIS} \label{sec:}

For extracting the strange quark PDF, the dimuon production data in
$\nu$-Fe DIS provide the most direct determination.  The basic channel is
the weak charged current process $\nu s \to \mu^- c X$ with a
subsequent charm decay $c \to \mu^+ X^{\prime}$.  These events provide
a direct probe of the $s W$-vertex, and hence the strange quark
PDF. In contrast, single muon production only provides indirect
information about $s(x,Q^2)$ which must then be extracted from a
linear combination of structure functions in the context of the QCD
parton model.  For this reason, fixed-target neutrino dimuon
production will provide a unique perspective on the strange quark
distribution of the nucleon in the foreseeable future.

The analysis of the recent NuTeV data is in progress.  This uses both
the  differential NLO calculation of the neutrino-induced DIS
charm production process and the Monte Carlo experimental detector
simulation program.  This project will extract the strange quark PDF
from the dimuon data at NLO with unprecedented accuracy.


In the long run, a high luminosity neutrino
factory could, of course, considerably raise the accuracy of present
day information from $\nu$-DIS, and allow us to investigate additional 
distributions. 
 For example, with high statistics  $\nu$-DIS data one can investigate
the character of Sudakov logarithms arising from multiple gluon
emission. The resummation of such soft gluons in the case of 
heavy quark production is a program that is currently under development. 
 High statistics  $\nu$-DIS data would allow us to not only study the
integrated distributions, but also study more differential distributions
which are sensitive to the details of the resummation procedure.

\head{$\Delta xF_3$} \label{sec:}

There are a number of outstanding puzzles where theoretical
expectations don't match experimental measurements, and $\Delta xF_3$
is a recent example.  $\Delta xF_3$ is obtained by taking the
difference of $F_3$ neutrino and anti-neutrino structure function; in
the simple LO parton model, this quantity is proportional to the
difference between the strange and charm quark PDF.
Specifically, 
$\Delta xF_3 
\simeq x F_3^{\nu N}-  x F_3^{\bar{\nu} N} 
\simeq 4 x\{ s-c\}$, where $N$ represent an isoscalar nucleon.

Theoretical predictions for $\Delta xF_3$ systematically undershoot
fixed target data at the $\sim 1\, \sigma$-level at low $x$ and $Q$.
The neutrino structure function $\Delta xF_3$ is obviously sensitive
to the strange sea of the nucleon and the details of deep inelastic
charm production.  A closer inspection reveals, however, considerable
dependence upon factors such as the charm mass, factorization scale,
higher twists, contributions from longitudinal $W^\pm$ polarization
states, nuclear shadowing, charge symmetry violation, and the PDF's.
This makes $\Delta xF_3(x,Q^2)$ an excellent tool to probe both
perturbative and non-perturbative QCD.

As the situation stands now, this $\Delta xF_3(x,Q^2)$ puzzle poses an
important challenge to our understanding of QCD and the related
nuclear processes in an important kinematic region. The resolution of
this puzzle is important for future data analysis, and the solution is
sure to be enlightening, and allow us to expand the applicable regime
of the QCD theory.\cite{Kretzer:2001mb}

\head{Heavy Quark Production} \label{sec:}


The production of heavy quarks, both hadroproduction and
leptoproduction, has become an important theoretical and
phenomenological issue.\cite{Frixione:1998ma}
 In part, heavy quark production has attracted much attention because
in many instances the theoretical expectations differ significantly
from experimental measurements.
 While the hadroproduction mode has a higher
mass reach ({\it e.g.}, the top-quark), the simpler leptoproduction
process can provide important insights into the fundamental production
mechanisms.

An important theoretical consideration which enters the calculations of
heavy quark production is the existence of multiple mass scales. For
example, in a DIS experiment the important scales are the virtuality
of the exchanged boson $Q$, and the mass of the heavy quark $M_Q$.
 This represents a significant theoretical challenge 
because extending theorems of  factorization and resummation 
to the case involving multiple scales is nontrivial.

Ideally we would like to explore the full kinematic range from 
$Q\ll M_Q$ as in heavy quark photoproduction, 
to the transition region $Q \sim M_Q$, 
and finally to $Q \gg M_Q$ where the mass effects become negligible.
 Setting the very massive top-quark aside, 
in the case of charm and bottom with masses in the few GeV range, 
fixed-target experiments are best suited to cover the kinematic range where
$Q\ll M_Q$  and  $Q \sim M_Q$; additionally
they can also cover a portion of the $Q \gg M_Q$ region.
 Therefore, a complete program to investigate heavy quarks must
consider information from a variety of sources including charm and
bottom production at both fixed-target and collider lepton and hadron
facilities.\cite{Gerber:1999xg}

One degree of freedom that has not been fully studied or exploited in
this area is the issue of an ``intrinsic" heavy quark component. While
the question of intrinsic heavy quarks has been discussed in the
literature for many years, it still remains unresolved.
 This is a controversial theoretical issue; hence a definitive
experiment is called for.  A particularly incisive test of this theory
would be to make precise measurements of heavy quark production in the
threshold region.  In this kinematic regime, the usual ``perturbative"
heavy quark component arising from gluon splitting ($g\rightarrow Q
\bar{Q}$) is comparatively small; therefore a measurement in this
region has more discriminating power to distinguish the ``intrinsic"
heavy quark component.
 Such an experiment could put the theoretical debate on ``intrinsic"
heavy quarks to rest once-and-for-all. Furthermore, such an experiment
will answer important questions regarding heavy quark production and
our ability to make accurate calculation in the presence of disparate
mass scales.

\head{Higher Twist} \label{sec:}

Higher twist (or power suppressed) corrections represent a
long-standing hurdle to making accurate theoretical predictions for
structure function data over the full kinematic range.  Higher twist
corrections should not simply be avoided; accurate characterization of
higher twist corrections provides new information on parton-parton
correlations within the nuclei.\cite{Kuhlmann:2000sf}

The kinematic limits where considerations of higher twist
contributions become important are 1) at high-$x$, and 2) at low $Q^2$
where terms of order $\Lambda^2/Q^2$ become significant.  In the
high-$x$ region, the limiting factor is primarily statistics.  In the
low $Q^2$ region the statistics are generally adequate, but if the
data is taken on heavy targets the higher twist effects are entangled
with nuclear effects.

Consequently, the ideal testing ground would be to have high
statistics measurements on a light nuclear target. This would allow
systematic separation of the higher twist effects from the nuclear
effects, and better allow us to learn about both in the process.

\head{Conclusions}

In recent years, new information has become available concerning
parton distributions and their uncertainties.  The issues have become
more important with the realization that these uncertainties could be
hampering searches for physics beyond the Standard Model.  
The
different topics reviewed in this report have clarified some of the
issues, but have also raised new questions to be addressed.  We have
outlined a program of measurements, as well as important theoretical
work, that is needed to improve the uncertainties in parton
distributions.
 Therefore, a complete program to investigate the nucleon structure
must consider information from a variety of sources including both
fixed-target and collider lepton and hadron facilities.

\head{Acknowledgment}

The author would like to thank 
B.~Fleming,
D.~Mason, 
W.~Melnitchouk,
J.~Morfin,
S.~Kretzer,
S.~Kuhlmann, 
P.~Reimer,
R.~Thorne, 
W.-K.~Tung,
and
U.-K.~Yang,
for 
helpful discussions.
 This research was supported
by the U.S. Department of Energy,
and by the Lightner-Sams Foundation. 




\end{document}